\documentclass[twocolumn,showpacs,amsmath,amssymb,groupedaddress,prl]{revtex4-1}

\usepackage{graphicx} 

\usepackage{xcolor}
\usepackage{color}      
\usepackage{soul} 

\definecolor{acolor}{rgb}{1,0.6275,0}
\definecolor{bcolor}{rgb}{0.4392,0.1882,0.6275}
\definecolor{column}{rgb}{0.0941,0.6039,0.9922}
\definecolor{line}{rgb}{0.8314,0,0.1137}
\definecolor{bluematlab}{rgb}{0,0.4470,0.7410}

\begin{document}

\title{Crackling dynamics in the mechanical response of knitted fabrics}

\author{Samuel Poincloux\textsuperscript{1}}
\email{spoincloux@lps.ens.fr}
\author{Mokhtar Adda-Bedia\textsuperscript{2}} 
\author{Fr\'ed\'eric Lechenault\textsuperscript{1}}
\affiliation{\textsuperscript{1}Laboratoire de Physique Statistique, Ecole Normale Sup\'erieure, PSL Research University, Sorbonne University, CNRS, F-75231 Paris, France}
\affiliation{\textsuperscript{2}Universit\'e de Lyon, Ecole Normale Sup\'erieure de Lyon, Universit\'e Claude Bernard, CNRS, Laboratoire de Physique, F-69342 Lyon, France}

\date{\today}
\begin{abstract}
Crackling noise, which occurs in a wide range of situations, is characterized by discrete events of various sizes, often correlated in the form of avalanches. We report experimental evidence that the mechanical response of knitted fabric displays such broadly distributed events both in the force signal and in the deformation field, with statistics analogous to that of earthquakes or soft amorphous materials. A knit consists of a regular network of frictional contacts, linked by the elasticity of the yarn. When deformed, the fabric displays spatially extended avalanche-like yielding events resulting from collective inter-yarn contact slips. We measure the size distribution of these avalanches, at the stitch level from the analysis of non-elastic displacement fields, and externally from force fluctuations. The two measurements yield consistent power law distributions reminiscent of those found in other avalanching systems. Our study shows that a knitted fabric is not only a thread-based metamaterial with highly sought after mechanical properties, but also an original, model system, with topologically protected structural order, where intermittent, scale-invariant response emerges from minimal ingredients, and thus a significant landmark in the study of out-of-equilibrium universality.
\end{abstract}

\maketitle

Crackling dynamics in materials mechanical response is currently intensively studied owing to its fundamental and industrial relevance and to the vast range of systems it encompasses. Indeed, examples of such response is usually found in disordered physical systems like granular materials~\cite{albert2000jamming,hayman2011granular,denisov2016universality}, foams~\cite{lauridsen2002shear}, metallic glasses~\cite{sun2010plasticity,antonaglia2014bulk}, seismic regions~\cite{chen1991self} or front propagation in heterogeneous media~\cite{bonamy2008crackling,chevalier2017avalanches}, but is also documented in structurally ordered situations~\cite{friedman2012statistics, sparks2018shapes, carrillo1998experimental}. In the case of soft amorphous materials, though the origin of elasticity and plasticity and their typical lengthscales~\cite{uhl2015universal} largely differ from one system to another, a common framework has been established to investigate and predict the avalanche features~\cite{bouchbinder2007athermal,dahmen2011simple,lin2014scaling}. 

\begin{figure}[htb]
	\includegraphics[width=0.9\linewidth]{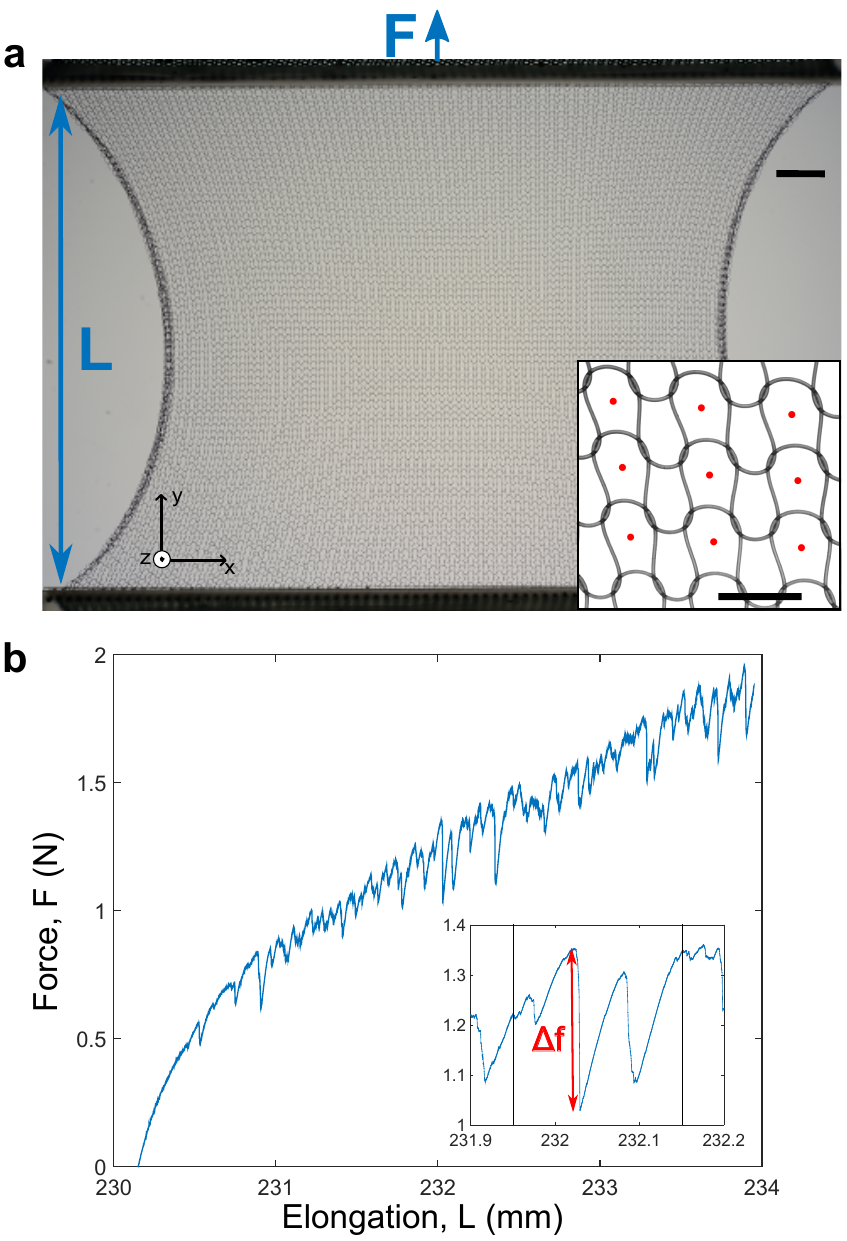}
\caption{Experimental system and typical force response. (a) A knitted fabric is stretched uniaxially while its mechanical response is recorded and the position of the stitches is tracked through digital image processing, with a precision of approximately $10\,\mu\mathrm{m}$. Typical picture of the knit; the stretching direction is materialized by the arrow associated to the force $F$ while $L$ denotes its elongation. The scale bar is $25\,\mathrm {mm}$ long. Inset: zoom over a few stitches tagged by a red dot indicating their position defined as their geometric center. Here the scale bar is $4\,\mathrm {mm}$ long. (b) Typical mechanical response of the fabric while stretched between $L_m=230\,\mathrm{mm}$ and $L_f=234\,\mathrm{mm}$ at a constant speed of $v=5\,\mu\mathrm{m}/s$. Stick-slip events at the contact points generate an intermittent signal, typical of crackling dynamics. Inset: zoom over a small interval, the definition of $\Delta f$ is emphasized and the two vertical lines distant by $\Delta L=0.2mm$ point to elongations at which two successive pictures of the fabric are taken.}
\label{fig:experiment}
\end{figure}

This work aims at demonstrating that despite its fundamentally ordered nature, knitted fabrics can also be studied within this framework. A knit is made of an elastic yarn, morphed into a 2D surface by imposing a topological, periodic pattern of self-crossing points, resulting in a network of so called stitches (Fig.~\ref{fig:experiment}a). Stitches deform elastically through bending of the yarn, but friction at the crossing points adds an uncertainty to the contact forces, inducing irreversible stick-slip activity. Those events propagate in the stitch network, generating avalanches and producing plastic events in the mechanical response. In this study, we use tools borrowed from the study of soft amorphous materials to characterize, externally and internally, the avalanches in this system, and illustrate why it provides a handy tool to make progress in this field.   
 
\textit{Experiments} -- We perform a tensile test on a model fabric, knitted out of a nylon monofilament of diameter $150\,\mu\mathrm{m}$ (Stroft$^{\mbox{\scriptsize{\textregistered}}}$ GTM), and record its stitch displacement fields and global mechanical response. The sample is crafted using a Toyota KS858 single bed knitting machine and is composed of $83\times83$ stitches with an average lateral and longitudinal size of respectively $3.9\,\mathrm {mm}$ and $2.8\,\mathrm{mm}$. It is then clamped on its upper and lower rows, preventing lateral displacement of the corresponding stitches. The tensile test consists on varying cyclically $L$, the elongation of the fabric along the so-called wale direction, between $L_{i}=215\,\mathrm{mm}$ and $L_f=234\,\mathrm{mm}$. The mechanical response is analyzed during the stretching phase on a shorter elongation range, between $L_m=230\,\mathrm{mm}$ and $L_f$. In this interval, the force needed to deform the fabric is recorded at high acquisition frequency with an Instron$^{\mbox{\scriptsize{\textregistered}}}$ (model 5965) mounted with a $50\,\textrm{N}$ load cell and pictures of the sample are taken every $\Delta L=0.2\,\mathrm{mm}$ increase in elongation. The pictures are taken at high resolution ($7360\times4912$ pixels) using a Nikon$^{\mbox{\scriptsize{\textregistered}}}$ D800 camera with a $60\,\textrm{mm}$ 1:2:8:G AFS MicroNikkor lens. To approach the quasi-static deformation limit in the interval $[L_m,L_f]$, we impose a constant pulling speed $v$ of the dynamometer and set it at a small value ranging from $1\,\mu\mathrm{m}/s$ to $10\,\mu\mathrm{m}/s$. To reduce the duration of the experiment, we fix $v=0.5\,\mathrm{mm}/s$ outside this measurement window. The imposed elongation $L$ as function of time is shown in Supplemental Fig.~S1~\cite{SupMat} and Supplemental Table~S1~\cite{SupMat} summarizes the parameters of all conducted tensile tests. Finally, a typical image of the fabric and the recorded force during one cycle are shown in Fig.~\ref{fig:experiment}.

\textit{Estimation of avalanche size} -- Upon stretching, the force signal displays, around an average elastic response, typical force fluctuations indicative of avalanches. The fluctuations consist in linear regions, stiffer than the average response, interrupted by plastic events provoking abrupt drops. The height of the drops $\Delta f$ can be measured and are expected to be correlated to  the avalanche size. Furthermore, evidences of those avalanches are identified in the deformation field of the stitch network. Performing an external measurement associated with an internal one is crucial to characterize the events and to rule out other possible phenomena.

\begin{figure*}[htb]
	\includegraphics[width=0.9\linewidth]{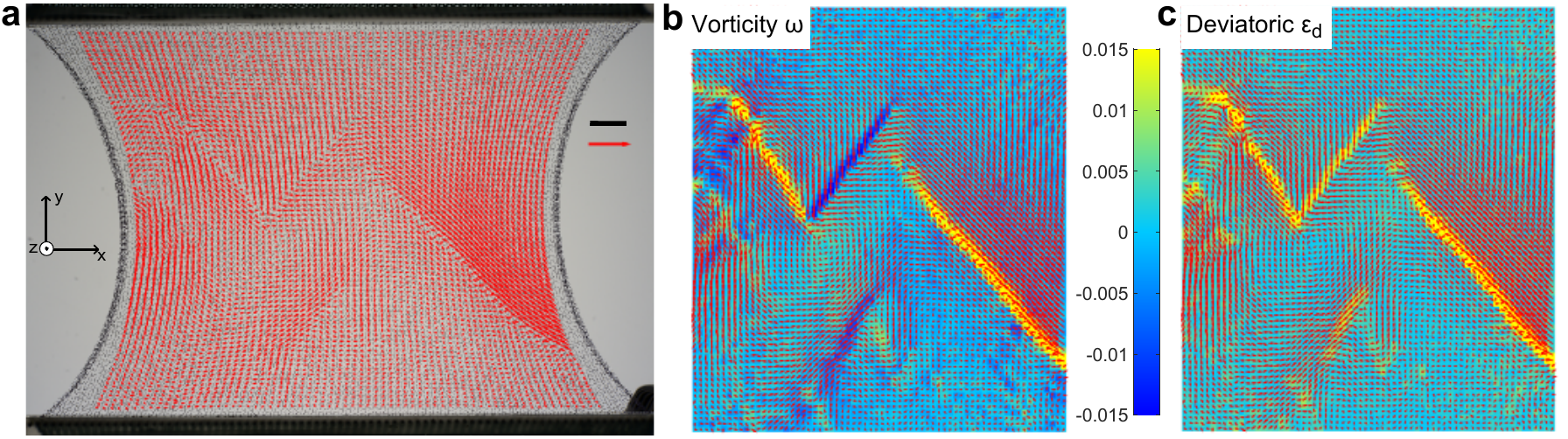}
\caption{Local detection of slip events. (a) Displacement field corresponding to the inset of Fig.~\ref{fig:experiment}b. Each stitch is tagged by its non-affine displacement $\vec u$, portrayed by red arrows magnified by a factor $35$. Black scale bar (position) $25\,\mathrm {mm}$, red scale bar (displacement)  $0.6\,\mathrm {mm}$. (b) vorticity $\omega$ and \textbf{c}, deviatoric strain $\varepsilon_d$ of the displacement field in the stitch network. Each arrow of the displacement field is associated to a single stitch.}
\label{fig:displacement}
\end{figure*}

Digital image processing allows to recover the position field of the stitch network and its displacement field between two successive pictures $\vec u _{tot}$ is computed. To emphasize its non-elastic component, the affine part $\vec u_{lin}$ is removed. We name $\vec u$ the resulting non-affine displacement field: $\vec u=\vec u _{tot}-\vec u_{lin}=u_x\vec e_x+u_y\vec e_y$, Supplemental Fig.~S2~\cite{SupMat} illustrates such operation. Fig.~\ref{fig:displacement}a shows that the non-affine displacement field appears highly heterogeneous, with abrupt spatial changes in the direction and size of $\vec u$ seemingly organized along diagonal lines. Those changes indicate that regions of the fabric are sliding against one another and are reminiscent of dislocation lines in crystals~\cite{papanikolaou2012quasi}. However, unlike the crystalline case, the connectivity of the network is locked and sliding events remain small compared to the size of a unit cell. On that account, in order to characterize the features in $\vec u$, we use two invariants of the deformation tensor~\cite{maloney2008evolution}, the vorticity $\omega=\frac{\partial u_y}{\partial x\hfill}-\frac{\partial u_x}{\partial y\hfill}$ and the deviatoric strain $\varepsilon_d=\sqrt{\left({\frac{\partial u_x}{\partial x\hfill}-\frac{\partial u_y}{\partial y\hfill}}\right)^2+\left({\frac{\partial u_y}{\partial x\hfill}+\frac{\partial u_x}{\partial y\hfill}}\right)^2}$.   The values of $\omega$ and $\varepsilon_d$ associated to the displacement field depicted in Fig.~\ref{fig:displacement}a are displayed on respectively Fig.~\ref{fig:displacement}b and Fig.~\ref{fig:displacement}c (See Movie~S1 in Supplemental Material~\cite{SupMat} showing these fields for different $\Delta L$ along the curve $F(L)$ of Fig.~\ref{fig:experiment}b). The boundaries between sliding regions of the knit are well captured by the two invariants $\omega$ and $\varepsilon_d$ which hence are good candidates to evaluate the size of the sliding events from the local measurements. In contrast, it is worth noticing that $\vec{\nabla}\cdot\vec{u}$ and the shear strain $\frac{\partial u_y}{\partial x\hfill}+\frac{\partial u_x}{\partial y\hfill}$ are always vanishingly small and show no significant variation in the vicinity of a sliding line (see Supplemental Fig.~S3~\cite{SupMat}). The sign of $\omega$ allows to discern two main event orientations: we define positive events those featuring $\omega>0$ and negative ones those with $\omega<0$. To retrieve an event size $S_\omega$ from the scalar fields $\omega$, we detect the connected stitches with $\left|\omega\right|$ higher than a threshold value and then integrate $\left|\omega\right|$ over those stitches. A demonstration of this process is shown in Supplemental Fig.~S4~\cite{SupMat}. The same operation is applied to measure the events size $S_{d}$ from $\varepsilon_d$.

Finally, we have verified that the location and size of sliding events are robust against the use of the total vector field $\vec u_{tot}$, instead of $\vec u$, for the definition of $\omega$ and $\epsilon_d$. This is mainly due to the fact that, even though a heterogeneous underlying loading is applied to the fabric, the spatial variations of $\vec u_{lin}$ are small compared to those due to plastic events.

\begin{figure}[htb]
	\includegraphics[width=0.9\linewidth]{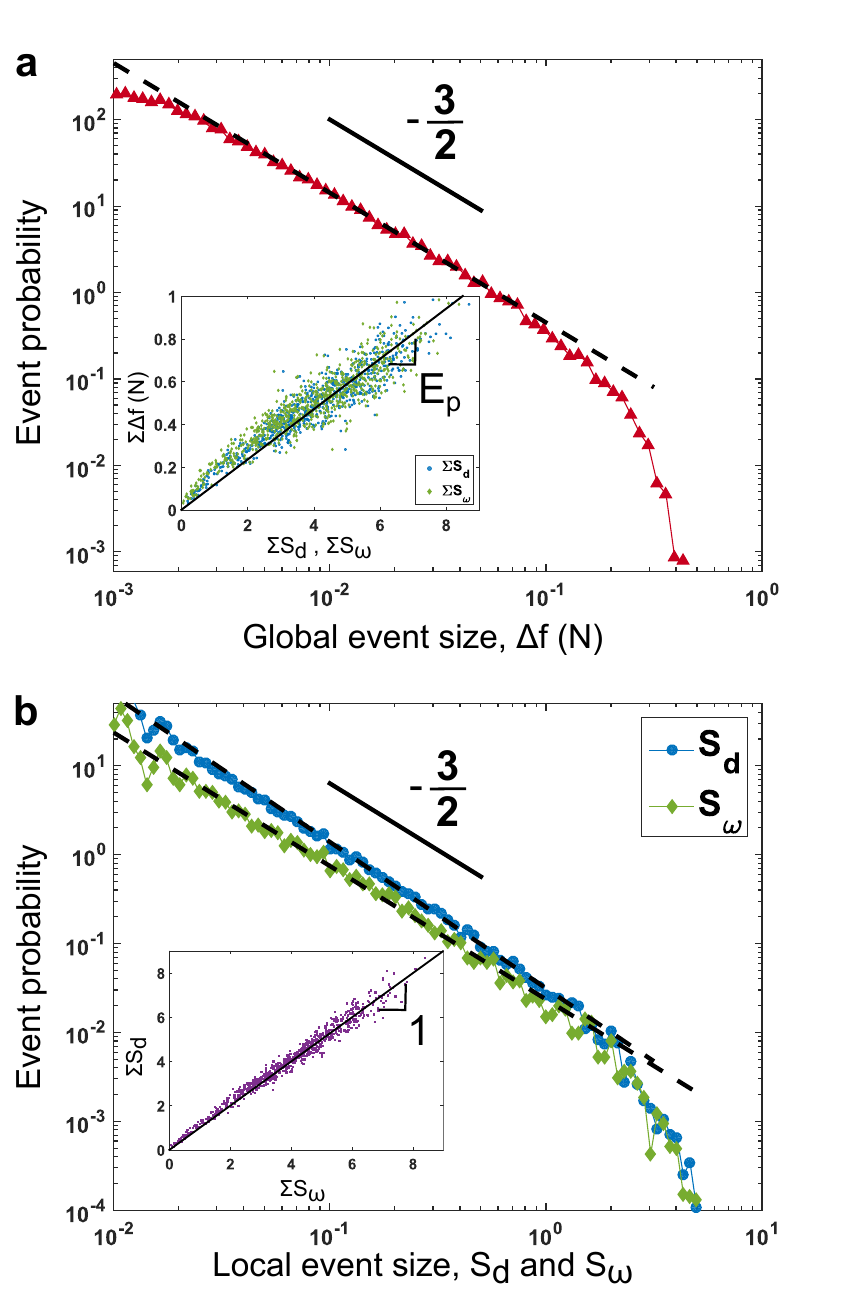}
\caption{Event size distribution measured from external and internal quantities. (a) Probability distribution of global event size measurement $\Delta f$. Dotted line is the best linear fit with a slope of $-1.50\pm0.03$. Inset: for each interval between two images, the sum of  event size measured externally is compared to the sum of events measured internally. (b) Probability distribution of local event size measurements from vorticity $S_\omega$ and deviatoric strain $S_{d}$ of the displacement field. Dotted lines are the best linear fit with a slope of $-1.51\pm 0.05$ and $-1.61\pm0.03$ for respectively $S_\omega$ and $S_{d}$. Inset: Comparison between the sum of $S_\omega$ and the sum of $S_{d}$ for each interval. The loading speed for the data shown in this figure is $v=5\,\mu\mathrm{m}/s$. The uncertainty in the exponents is evaluated from standard error and a $95\%$ confidence interval.}
\label{fig:distribution}
\end{figure}

\textit{Distribution of avalanche size} -- We now have a tool to measure the size of 'quake-like' events, from an external perspective with the force drops $\Delta f$, but also internally using two different means: high vorticity regions $S_\omega$ and high deviatoric strain regions $S_{d}$.  The protocol, and especially $\Delta L$, is chosen such that the interval between pictures is much longer than the duration of an event, hence, each image is not correlated to the previous one, and each cycle can be seen as another, statistically independant trial. In that way, we can build up statistics to characterize the probability distribution of event size. Fig.~\ref{fig:distribution}a shows this distribution for $\Delta f$ while  Fig.~\ref{fig:distribution}b shows the ones for $S_\omega$ and $S_{d}$. The three distributions exhibit a power law decay with exponents of $-1.50\pm0.03$ for $\Delta f$, $-1.61\pm 0.03$ for $S_d$ and $-1.51\pm 0.05$ for $S_\omega$. Those power law distributions are characteristic of avalanching phenomena~\cite{sun2010plasticity,antonaglia2014bulk,bonamy2008crackling} and the exponents we find are consistent with the prediction $-3/2$ of mean field models~\cite{dahmen2011simple} for soft amorphous solids. However, the universality of this exponent is still debated~\cite{lin2014scaling,liu2016driving,denisov2016universality}. These scaling laws are robust upon varying the threshold value of $\left|\omega\right|$ and $\varepsilon_d$ for the event detection (Supplemental Fig.~S5), the loading speed (Supplemental Fig.~S6 for $\Delta f$ and Fig.~S7 for $S_\omega$ and $S_d$) or the stretching range (Supplemental Fig.~S8)~\cite{SupMat}. Internal and external measurements of event size have noticeably similar distribution, so one should probe if they are indeed two aspects of the same avalanches~\cite{amon2012hot,bares2017local}. Thus, for each interval between two images, we sum $S_\omega$ and $S_{d}$ over all the events detected within, giving respectively $\Sigma S_\omega $ and $\Sigma S_d$, and compare them to the sum of $\Delta f$, named $\Sigma \Delta f$, measured during the same interval. The resulting scatter plot (Fig.~\ref{fig:distribution}a, inset) shows a clear linear tendency, which establishes a statistical correspondence between internal and external events. The slope $E_p=0.12\,\mathrm{N}$ allows to extract an avalanche parameter relating plastic deformation and force drops. Besides, comparing $\Sigma S_\omega$ and $\Sigma S_{d}$  ( Fig.~\ref{fig:distribution}b, inset) reveals proportionality with a coefficient close to $1$. This suggests that in this system, the sliding events are also characterized by a strong correlation between the deviatoric strain and the vorticity of the displacement field, as retrieved theoretically below.

\textit{Avalanche propagation} -- Though it does not flow, the system at hand is reminiscent of soft amorphous solids which are commonly described using elasto-plastic models~\cite{nicolas2017deformation}. These approaches assume the material as an elastic matrix with a plastic (or yield) limit, and a distance to this limit distributed inhomogeneoulsy in space~\cite{baret2002extremal,agoritsas2015relevance}. While the stress is globally increased in the material, areas close to plastic limit will yield first and induce a stress redistribution that may trigger other plastic events~\cite{desmond2015measurement,cao2018soft}, resulting in propagating avalanches~\cite{maloney2006amorphous,tanguy2006plastic,bouchbinder2007athermal,tyukodi2016depinning}. To test if a knitted fabric fits in this framework, we first analyze  the nucleation and morphology of plastic events. The viewing of different images shows that the events can actually intersect, although a V-shaped morphology seems to be the generic feature. To further assess this specific feature, we performed high-speed imagery of an avalanche (see Fig.~S9 in \cite{SupMat}). It turns out that avalanches often start from a single, bulk stitch, and then propagate from this particular site in all possible favoured directions.

\begin{figure}[htb]
	\includegraphics[width=0.9\linewidth]{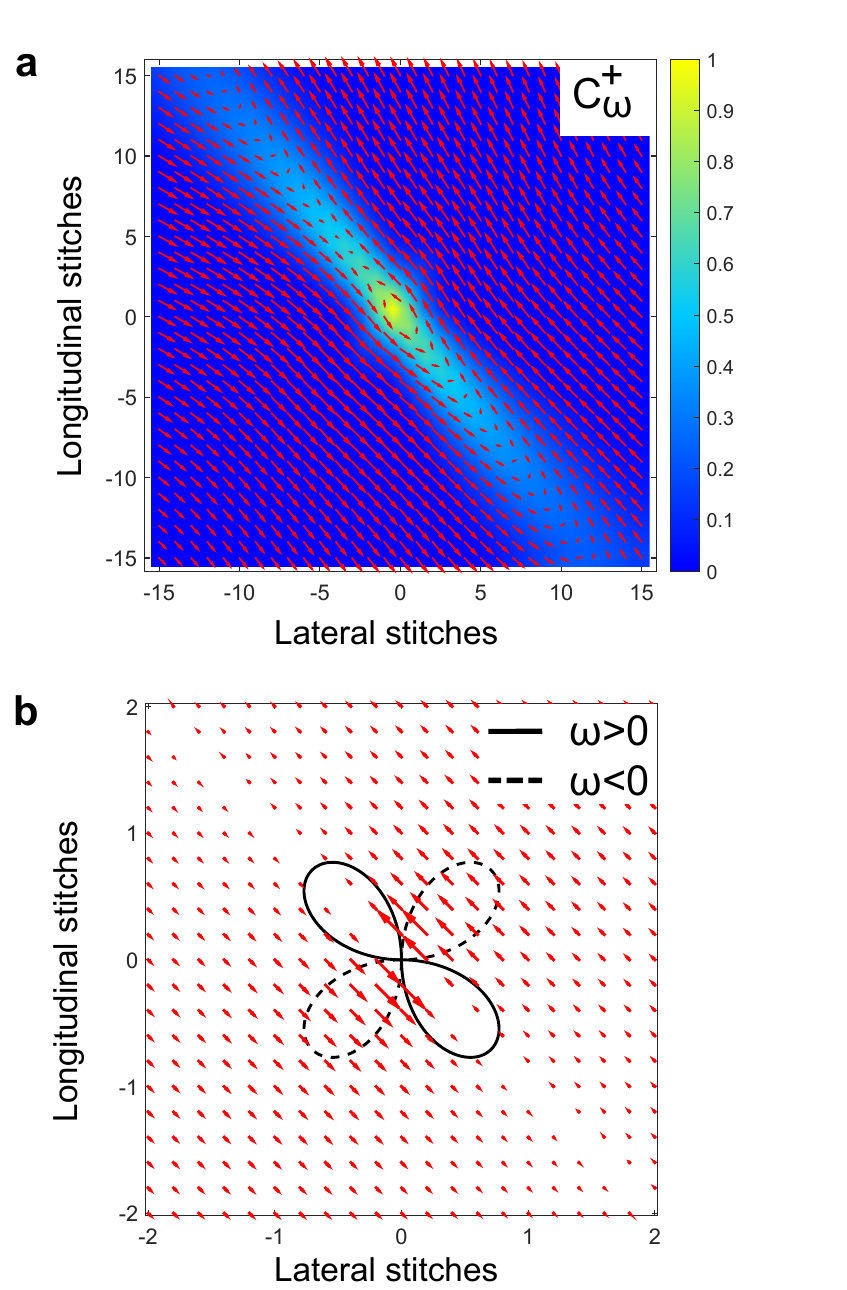}
\caption{Experimental spatial correlation and theoretical elastic deformation of positive events. (a) Amplitude of the correlation function $C^+_{\omega}$ of the vorticity, showing the events propagating along diagonal stitches, together with the relative displacement field during positive events $\vec{\tilde{u}}_{+}(\delta_x,\delta_y)=\langle \vec u (x+\delta x,y+\delta_y)-\vec u(x,y)\rangle_+$. Data shown for  for $v=5\,\mu\mathrm{m}/s$. (b) Computed response of the elastic network to a local perturbation shown through the displacement field, along with the radial amplitude of the vorticity field.}
\label{fig:correlation}
\end{figure}

Now, let us study how plastic events are correlated in space~\cite{le2014emergence}. Since high values of the vorticity in the non-affine displacement field $\omega$ are a good signature of plastic events in our fabric, events spatial correlation can be evaluated with the following quantity: 
\begin{equation}
C^{\pm}_{\omega}(\delta x,\delta y)=1+\frac{\langle \omega(x+\delta x,y+\delta y)-\omega(x,y)\rangle_{\pm}}{\langle\omega(x,y)\rangle_{\pm}}
\end{equation}
where the average $\langle\rangle_{\pm}$ runs over all the stitches detected in a positive $(+)$ or negative $(-)$ event. $C^{+}_{\omega}$, displayed in Fig.~\ref{fig:correlation}a, presents a strong correlation in the direction $-\frac{\pi}{4}$ indicating that positive events propagate along the diagonal of the stitch network. For a negative event the result is the same but with the direction $\frac{\pi}{4}$. To uncover the relation between the avalanche propagation and how the elastic matrix reacts to a local plastic event, we use a framework~\cite{poincloux2018geometry} which provides with a continuous model of knit elasticity. Considering a homogeneous fabric, we locally impose a non-zero vorticity $\omega_0$ and deviatoric strain $\varepsilon_{d_0}$ over a region of size $d$,  with $\omega_0>0$ for a positive event and $\omega_0<0$ for a negative event, while $\varepsilon_{d_0}<0$ for both type of events. We then compute the resulting displacement field with vanishing displacement far from the perturbation. In the stitch network, the vorticity and deviatoric strain have the following expressions in polar coordinates $(r,\theta)$: $\omega (r,\theta)=\frac{\varepsilon_{d_0} d^2}{2  r^2}\sin{2\theta}$ and $\varepsilon_d (r,\theta)=\frac{\omega_0 d^2}{2 r^2}\sin{2\theta}$, valid for $r\geq d$. More details on the elastic model and calculations can be found in Supplemental Material~\cite{SupMat}. The resulting displacement field around a positive event is shown in Fig.~\ref{fig:correlation}b, together with the angular variation of $\omega (r,\theta)$. The elastic response of the knit allows to retrieve two properties of the measured events. First, the maxima of the vorticity and deviatoric fields are located along the same directions as those measured experimentally, irrespective of the event sign. Secondly, the  elastic model gives $\omega (r,\theta)$ directly proportional to $\varepsilon_{d_0}$, along with $\varepsilon_d (r,\theta)$ proportional to $\omega_0$, suggesting that during event propagation, vorticity and deviatoric are strongly correlated as evidenced experimentally ( see inset of Fig.~\ref{fig:distribution}b).

\textit{Conclusion} -- In this study, we show that the mechanical response of knitted fabric displays crackling dynamics in its mechanical response through stick-slips events, despite its topologically protected structural order, thus it is not prone to either structural rearrangement or yielding/failure. Global and local avalanche size display power law distributions as those predicted by mean-field models of soft amorphous materials. This approach differs from previous studies on friction in textile~\cite{matsuo2009hysteresis,dusserre2015modelling} and may trigger new perspectives for the study of textile mechanics. Moreover, the quality of the experimental measurements of the avalanches statistics in this original system ends up rivaling with the latest similar experimental analysis on more commonly studied systems~\cite{antonaglia2014bulk,denisov2016universality,bares2017local}. Knitted fabric can thus be used as a tool to investigate universal crackling response, allowing to distinguish between the effects of plastic threshold distributions present here and the missing structural disorder. This approach also proves advantageous for several reasons such as a straightforward experimental implementation and analysis, or the presence of numerous easily tunable parameters. 

\textit{Acknowledgments} -- The authors thank E. Agoritsas, A. Rosso, J.-L. Barrat for fruitful discussions. This work was carried out in the framework of the METAMAT project  ANR-14-CE07-0031 funded by Agence Nationale pour la Recherche. 

%

\end{document}


\centerline{\bf \large Supplemental Material for:}
\centerline{\bf \large Crackling dynamics in the mechanical response of knitted fabrics}

\section*{Profile of the tensile test}

\begin{figure*}[h]
\centerline{
	\includegraphics{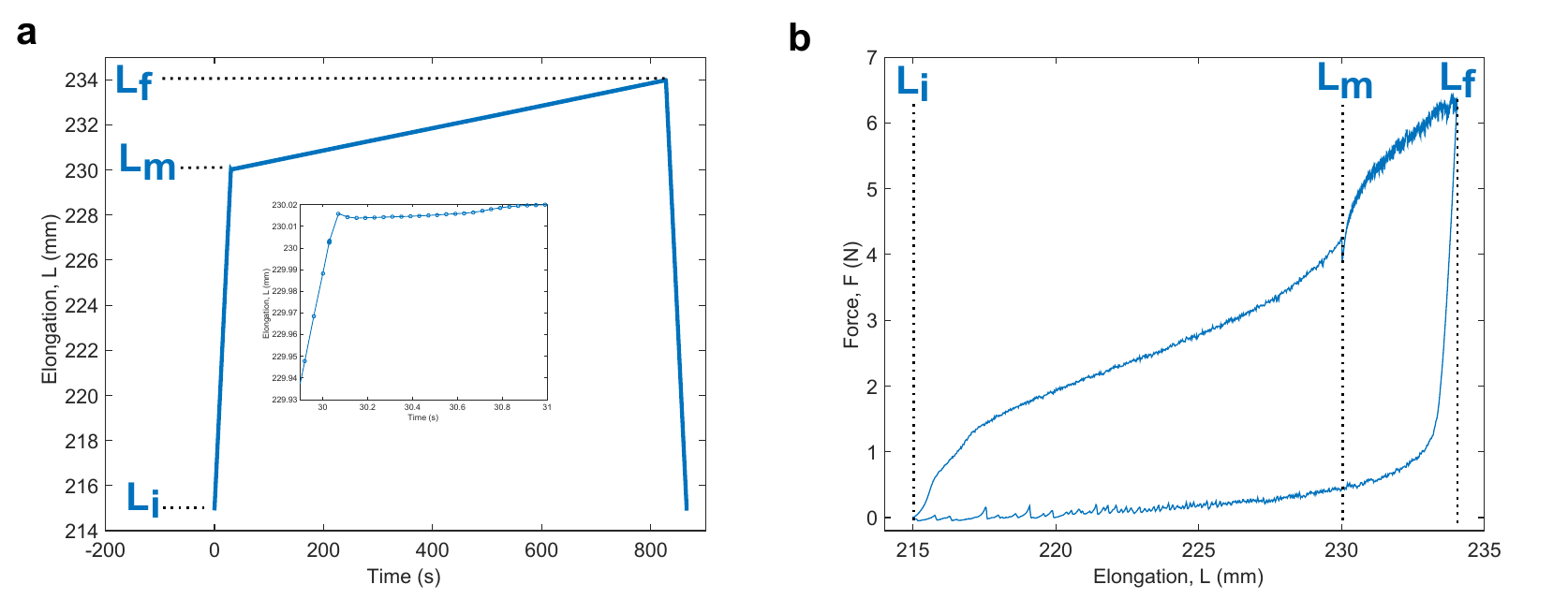}
}
\caption{\textbf{a}, Elongation of the fabric $L$ as a function of time for one cycle and $v=5\,\mu\mathrm{m}/s$ in the measurement range $[L_m,L_f]$. Inset: Zoom over the speed change around $L_m$. Due to the inertia of the motor, $L$ slightly decreases before adopting the speed $v=5\,\mu\mathrm{m}/s$. \textbf{b}, Mechanical response of the fabric during the whole cycle shown in \textbf{a}. At $L=L_m$, the force drops because of the slight decrease of $L$ combined with creep relaxation of the system. However, this effect nor the value of the measurement range $[L_m,L_f]$ don't appear to have a significant effect on the distribution of plastic events since we recover the same power law distribution when a constant speed of  $v=5\,\mu\mathrm{m}/s$ is maintained between $L_i$ and $L_f$ and the measurement range is varied (see Fig. S8).}
\label{fig:non_affine}
\end{figure*}

\newpage

\section*{Measurements of the non-affine displacement field}

\begin{figure*}[h]
\centerline{
	\includegraphics{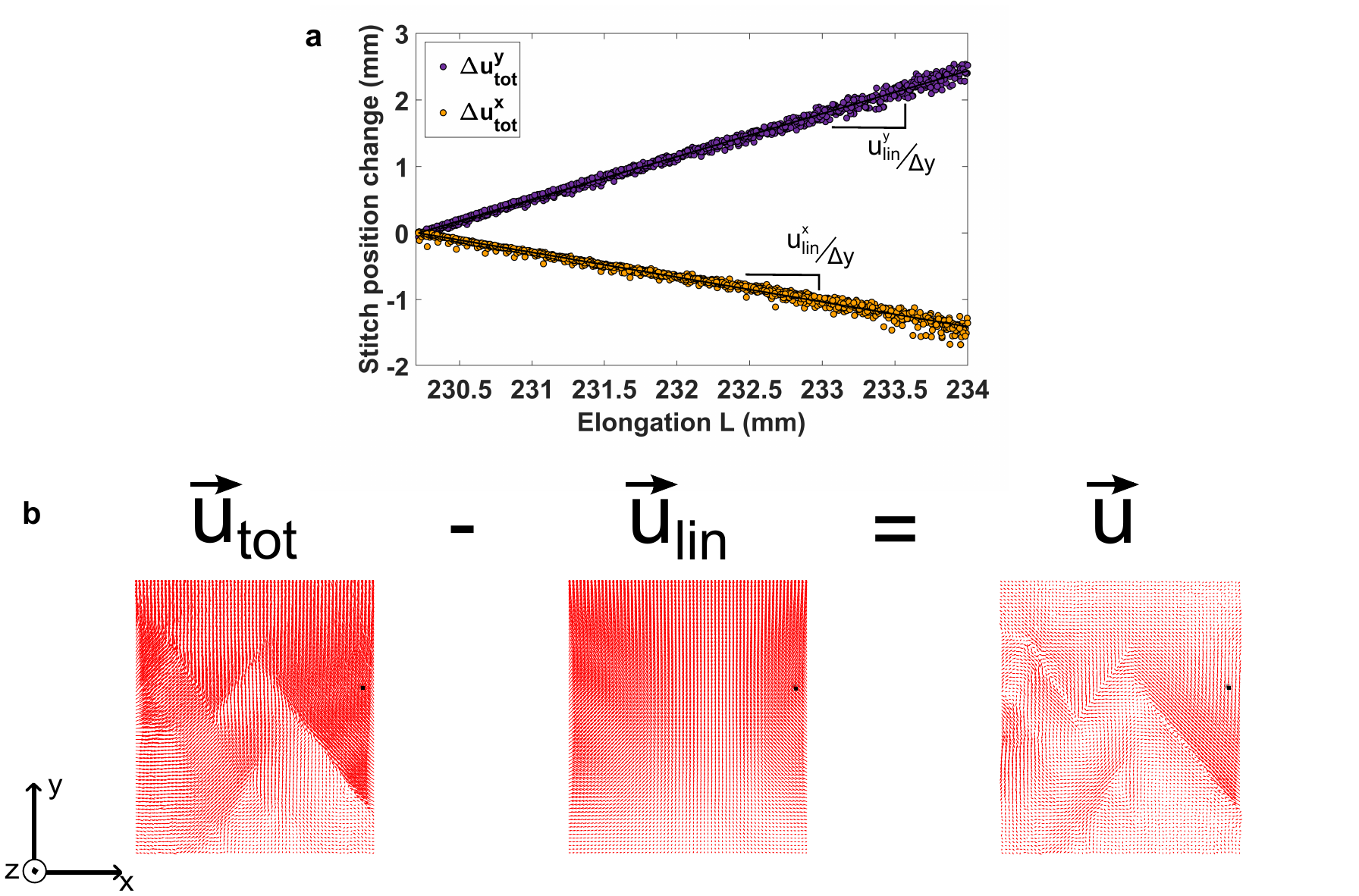}
}
\caption{\textbf{a}, Trajectory of a given stitch while the fabric is stretched from $L=230\,\mathrm{mm}$ to $234\,\mathrm{mm}$ (one point is one picture, all the cycles are displayed). On average, the stitches follow a linear trajectory that we can remove to determine the non-affine displacement field. \textbf{b}, Example of the total displacement field $\vec u_{tot}$, the linear displacement field $\vec u_{lin}$ estimated for each stitch from the slopes of its trajectory and the resulting non-affine displacement field $\vec u$. The black square locates the stitch considered in \textbf{a}.}
\label{fig:non_affine}
\end{figure*}

\newpage

\section*{Quantification of the non-affine displacement field spatial variations} 

\begin{figure*}[h]
\centerline{
	\includegraphics[height=0.75\textwidth]{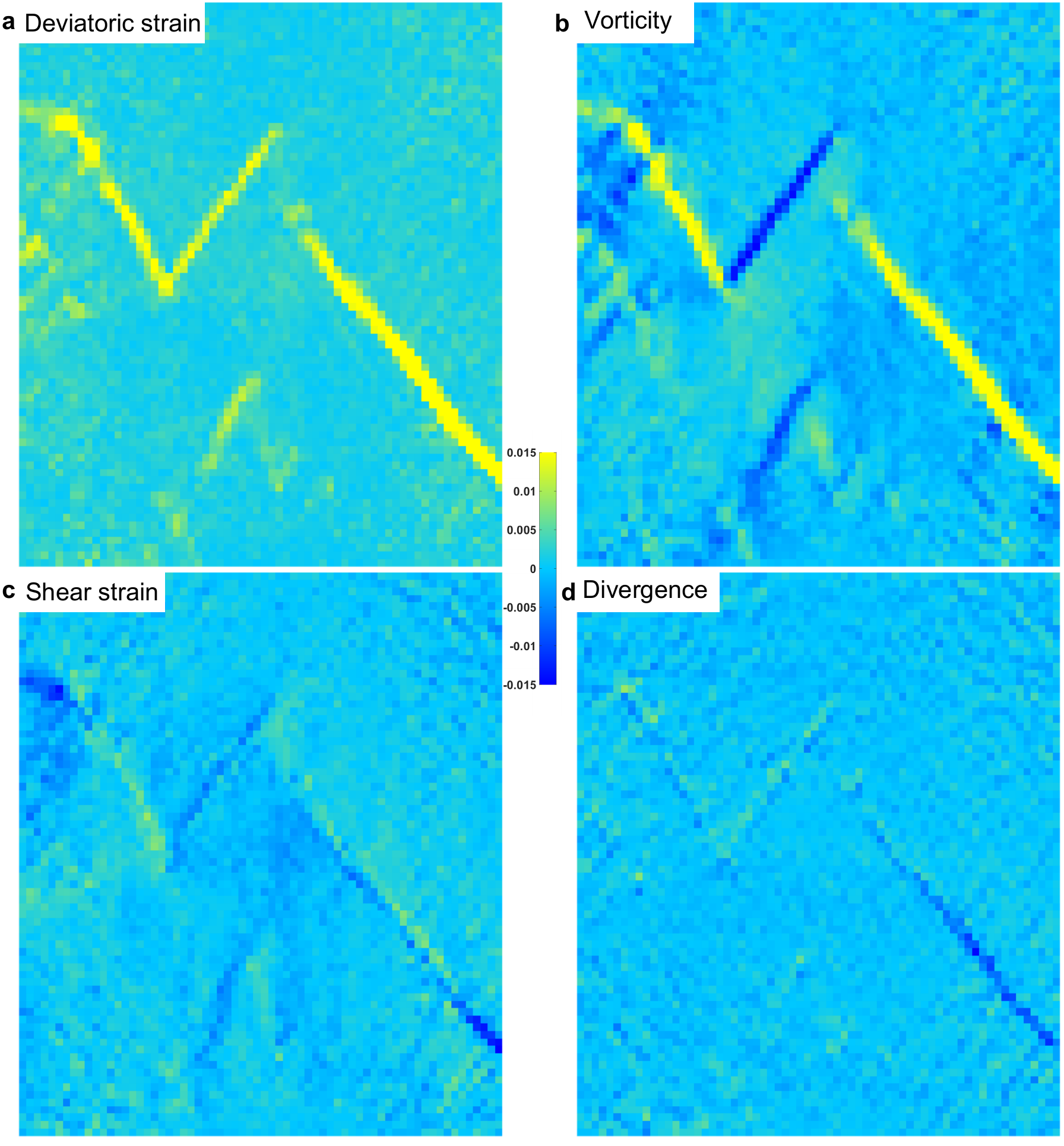}
}
\caption{\textbf{a}, Deviatoric strain defined by $\varepsilon_d=\sqrt{\left({\frac{\partial u_x}{\partial x\hfill}-\frac{\partial u_y}{\partial y\hfill}}\right)^2+\left({\frac{\partial u_y}{\partial x\hfill}+\frac{\partial u_x}{\partial y\hfill}}\right)^2}$. \textbf{b}, Vorticity defined by $\omega=\frac{\partial u_y}{\partial x\hfill}-\frac{\partial u_x}{\partial y\hfill}$. \textbf{c}, Shear strain defined by $\frac{\partial u_y}{\partial x\hfill}+\frac{\partial u_x}{\partial y\hfill}$. \textbf{d}, Divergence of $\vec{\nabla}\cdot\vec{u}$.}
\label{fig:div_shear}
\end{figure*}

\newpage

\section*{Measurements of the size of the events from the vorticity field} 

\begin{figure*}[h]
\centerline{
	\includegraphics{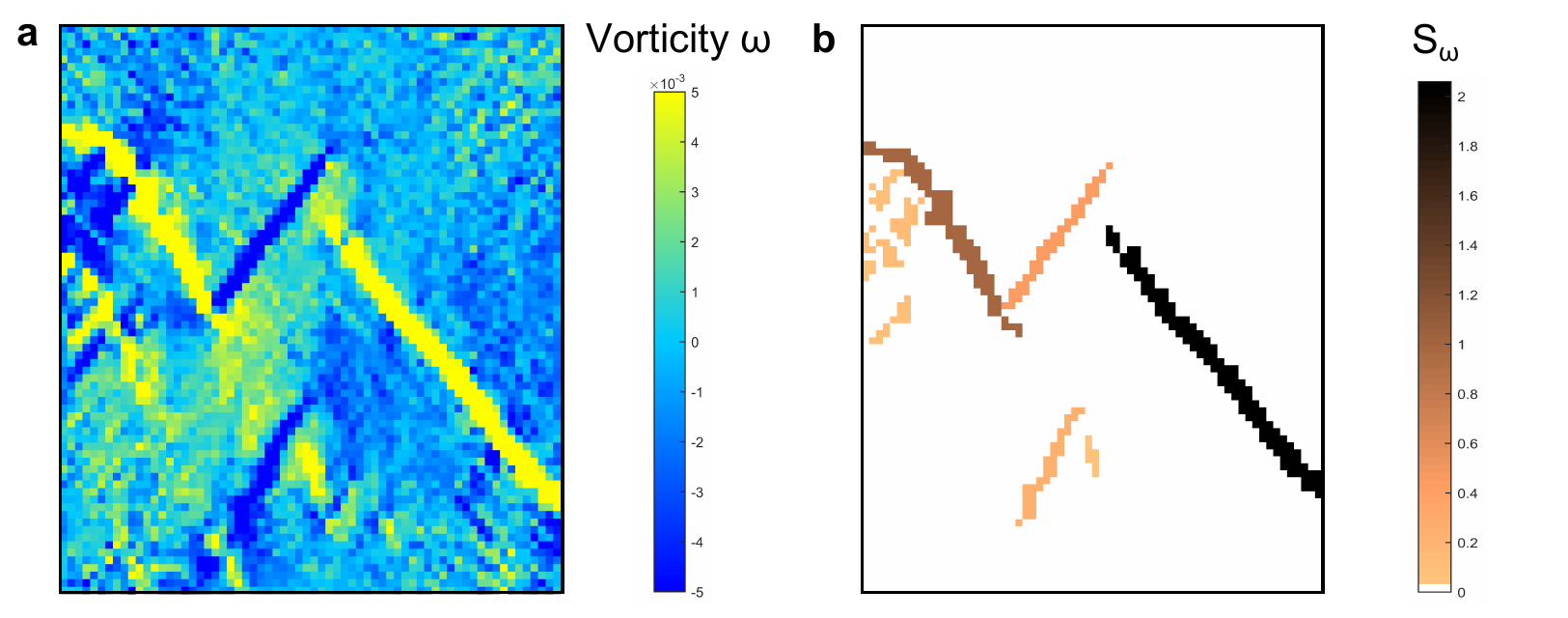}
}
\caption{\textbf{a}, Vorticity field $\omega$ with the color-scale saturated at the threshold value $\pm0.005$, one pixel of the image is one stitch. Every connected stitches with $|\omega|>0.005$ are considered to belong to the same event. \textbf{b}, Illustration of the events detected in the image in \textbf{a}. Each event is colored by its size $S_\omega$. Events connected but having an opposite sign of $\omega$ are not considered to belong to the same avalanche.}
\label{fig:disp_e&vent}
\end{figure*}

\newpage

\section*{Variability of the event size distribution with the threshold value} 

\begin{figure*}[h]
\centerline{
	\includegraphics{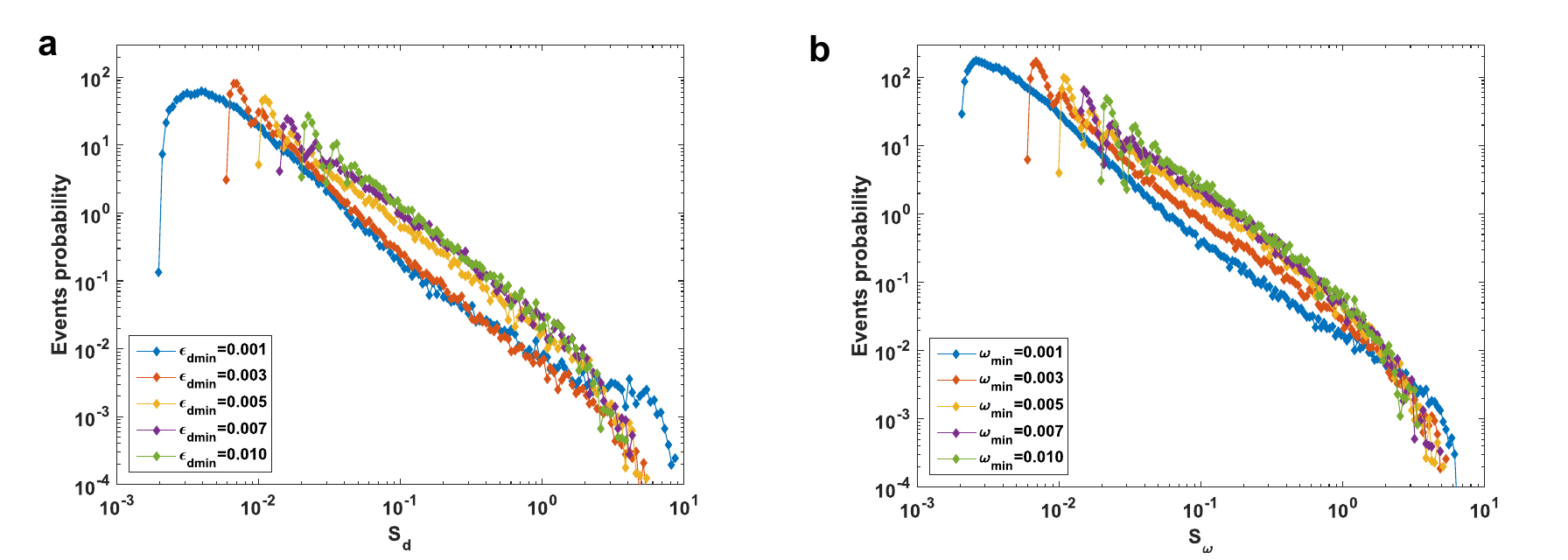}
}
\caption{\textbf{a}, Probability distribution of $S_d$ depending on the threshold value of $\varepsilon_d$ above which a stitch is considered to belong to an event. For small threshold, the density of large events are larger because it connects events that are considered distinct for higher value of the threshold. The power law distribution remains the same as soon as the bump at large event size disappears. \textbf{b}, Probability distribution of $S_\omega$ depending on the threshold value of $\varepsilon_d$ above which a stitch is considered to belong to an event. It shows the same behavior as $S_d$. For both $\omega$ and $\varepsilon_d$, the threshold value finally chosen is $0.005$, for all the speed.}
\label{fig:threshold}
\end{figure*}

\section*{Experimental parameters}
\begin{table}[h]
\begin{tabular}{|c|c|c|c|c|c|c|}
  \hline
   \multirow{2}{*}{Speed ($\mu m/s$)} & \multirow{2}{*}{Force sampling Frequency  (Hz)} & \multirow{2}{*}{$\Delta L$  (mm)} &  \multirow{2}{*}{Number of cycles} &\multicolumn{3}{|c|}{Number of events} \\ & & & & $\Delta_f$ & $S_d$ & $S_\omega$ \\
  \hline
1 & 25 & 0.2 & 10 & 5630 & 4488 & 3873\\
\hline
3 & 14 & 0.2 & 50 & 6722 & 14401 & 10880 \\
\hline
5 & 25 &0.2 &50 & 30395 &25884 & 21122 \\
\hline
8 & 20 & 0.2 & 80 & 15525 & 31156 & 23948 \\
\hline
10 & 25 &0.2 &80 & 31323 &54786 &44169 \\
\hline
\end{tabular}
\caption{Summary of the experimental parameters (speed, frequency, $\Delta L$, number of cycles) and the number of detected events ($\Delta_f$, $S_\omega$ and $S_{d}$) with the different methods.}
\end{table}

\newpage

\section*{Influence of pulling speed on $\mathbf{\Delta_f}$}

\begin{figure*}[h]
\centerline{
	\includegraphics{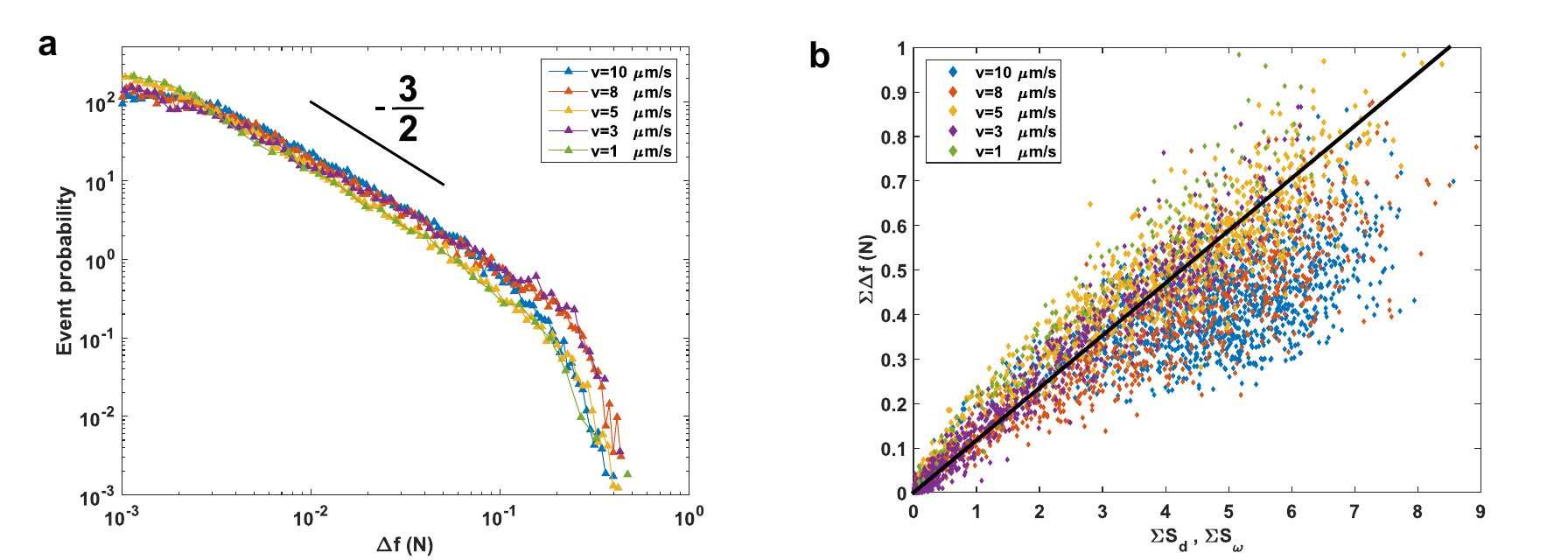}
}
\caption{ \textbf{a}, Probability distribution of $\Delta_f$ for the five tested speeds: no significant difference is observed. \textbf{b}, Correlation between the intensity of events per picture measured with $\Delta_f$ and measured with the displacement field. Even if the general tendency is conserved, increasing the speed tends to lower the coefficient of proportionality and also to decorrelate more and more the two measures.}
\label{fig:speed_force}
\end{figure*}

\newpage

\section*{Influence of pulling speed on $\mathbf{S_d}$ and $\mathbf{S_\omega}$}

\begin{figure*}[h]
\centerline{
	\includegraphics{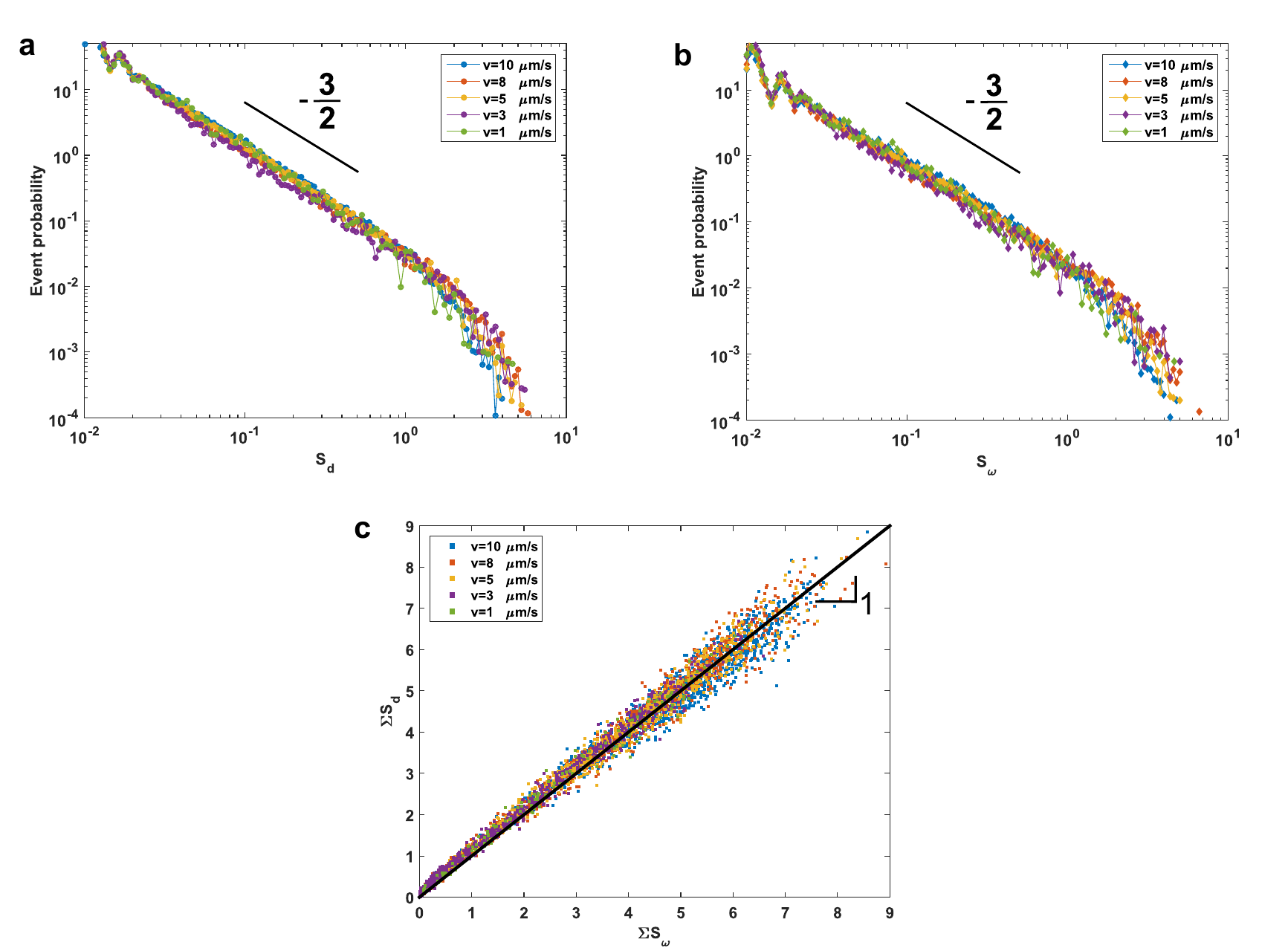}
}
\caption{ \textbf{a},  \textbf{b}, Probability distribution of respectively the event size measured with the deviatoric strain $S_d$ and with the vorticity $S_\omega$ for the five tested speeds: for both, no significant difference is observed. \textbf{c}, Correlation between the sum of $S_d$ and $S_\omega$ for all the images and all the speeds. A correlation with a coefficient of proportionality of $1$ is measured for all the different speeds.}
\label{fig:speed_disp}
\end{figure*}

\newpage

\section*{Influence of elongation range on the distribution of $\mathbf{S_\omega}$}

\begin{figure*}[h]
\centerline{
	\includegraphics{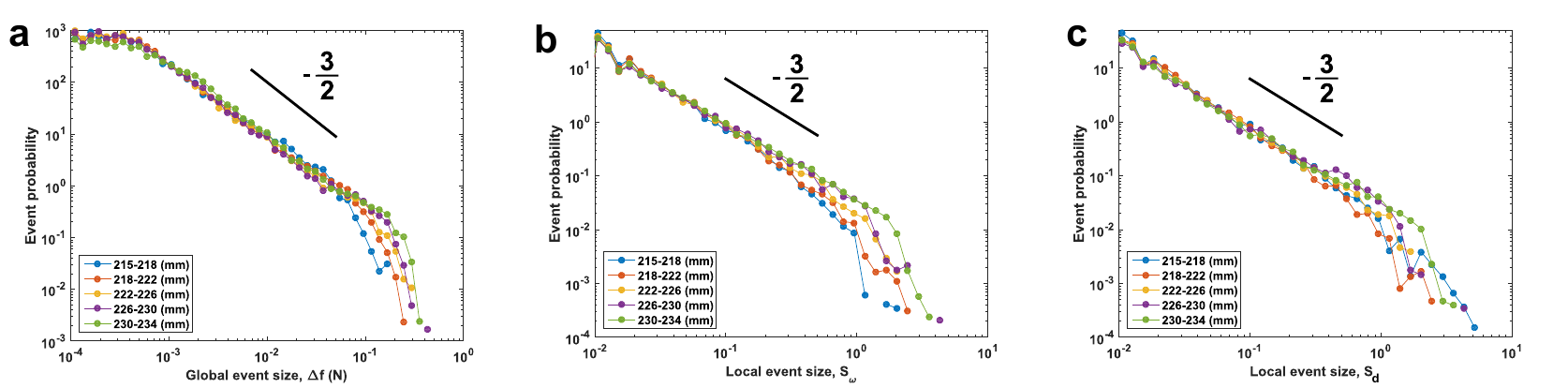}
}
\caption{The fabric is stretched at $v=5\,\mu\mathrm{m}/s$ between $L_i=215\,\mathrm{mm}$ and $L_f=230\,\mathrm{mm}$ and the probability distribution of $\Delta f$ (\textbf{a}), $S_\omega$ (\textbf{b}) and $S_d$ (\textbf{c}) are plotted for five different elongation ranges $L_m-L_f$. The power law of the distribution is not affected by the elongation range but the cut-off at large event size shows the tendency to grow as elongation increases. This indicates that as the global stress in the fabric increases with elongation, it allows to trigger larger events.}
\label{fig:speed_disp}
\end{figure*}

\newpage

\section*{Nucleation of plastic events}

\begin{figure}[h]
\begin{centering}
      \includegraphics[width=0.9\linewidth]{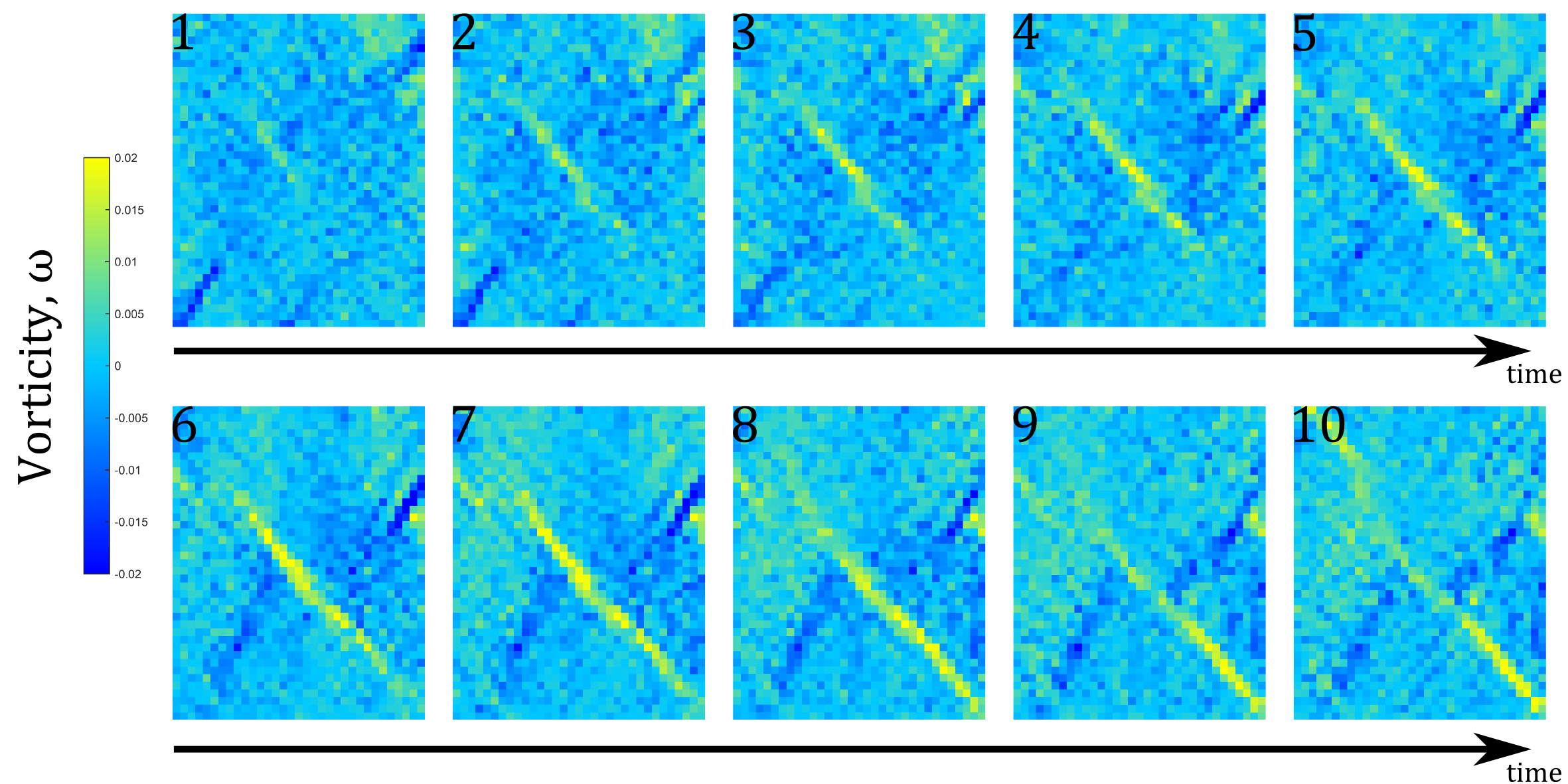}
   \caption{Birth and propagation of a plastic event as seen in the vorticity field with high-speed imaging: the event starts  in the bulk and propagates downwards. The loading velocity is $0.75$mm/s, the vorticity is computed between images distant of $0.2$mm as in the main text and images are separated by $40 \mu$s or equivalently $30 \mu$m.}
   \label{fig:fig2}
      \end{centering}
\end{figure}

\newpage

\section*{Elastic response of a knitted fabric to a local perturbation}

We use the formalism introduced in a previous paper~\cite{poincloux2018geometry}. We assume an isolated infinite homogeneous network, of periodicity $c^*\vec e_x$ along the \textit{course} (line) direction and $w^*\vec e_y$ along the wale (column) direction. A perturbation is introduced in $(x=0,y=0)$ and the resulting displacement field $\vec u$ is calculated while assuming no deformation far from the center. The size of the unit cells $\vec c$ and $\vec w$ are hence expressed as:
\begin{align}
\vec c (x,y)&=c^*\left({\vec e_x+\frac{\partial \vec u}{\partial x \hfill}}\right) \\
\vec w (x,y)&=w^*\left({\vec e_y+\frac{\partial \vec u}{\partial y \hfill}}\right)
\end{align}

The Lagrangian of the fabric reads:
\begin{align}
\mathcal{L}\{\protect\vec{c},\protect\vec{w}\}=&\tilde Y{\iint\limits_{x,y}\mathrm{d}x\mathrm{d}y\left({\frac{1}{c}+\frac{\beta}{w}}\right)+\alpha \iint\limits_{x,y}\mathrm{d}x\mathrm{d}y\left({c+\delta w}\right) } \nonumber\\ 
&-\iint\limits_{x,y}\mathrm{d}x\mathrm{d}y\,\vec{T}(x,y).\left({\frac{1}{c^*}\frac{\partial\vec{c}}{\partial y}-\frac{1}{w^*}\frac{\partial\vec{w}}{\partial x}}\right) \nonumber\\  
\end{align}
The first term is the elastic energy evaluated from the bending of the yarn, the second term is a constraint imposing the conservation of the yarn length and the third term enforces the local topological constraint; $\alpha$ and $\vec{T}(x,y)$ being their respective Lagrange multiplier. $\tilde Y$ is an effective stretching modulus, $\delta$ a parameter embedding the geometry of the stitch and $\beta$ a coefficient transcribing the asymmetric energy contributions: if $\vec c =c^*\vec e_x$ and $\vec w =w^*\vec e_y$ is the rest state, we get $\beta=\delta\left({\frac{w^*}{c^*}}\right)^2$. Variation of the Lagrangian with respect to $\vec c$ : $\mathcal{L}\{\vec c+\vec{\delta c},\vec w\}- \mathcal{L}\{\vec c,\vec w\}=0$, gives the following equation and its corresponding boundary condition.
\begin{align}
-\tilde Y\frac{\vec{c}}{c^3}+\alpha\frac{\vec{c}}{c}+\frac{1}{c^*}\frac{\partial\vec{T}}{\partial y}&=\vec 0 \label{eq:C}\\
\left[{\vec{T}.\vec{\delta C}}\right]_{y=\pm\infty}&=0
\end{align}
And with respect to $\vec w$ : $\mathcal{L}\{\vec c,\vec w+\vec{\delta w}\}- \mathcal{L}\{\vec c,\vec w\}=0$
\begin{align}
-\tilde Y \beta\frac{\vec{w}}{w^3}+\delta\alpha\frac{\vec{w}}{w}-\frac{1}{w^*}\frac{\partial\vec{T}}{\partial x}&=\vec 0 \label{eq:W} \\
\left[{\vec{T}.\vec{\delta W}}\right]_{x=\pm\infty}&=0
\end{align}
 The boundary conditions are satisfied whatever the value of $\vec T$ because we assume no deformation far from the center. The two previous equations can be combined to eliminate the Lagrange multiplier $\vec T$:
\begin{equation}
c^*\frac{\partial}{\partial x}\left[{-\tilde Y \frac{\vec{c}}{c^3}+\alpha\frac{\vec{c}}{c}}\right] +w^*\frac{\partial}{\partial y}\left[{-\tilde Y \beta\frac{\vec{w}}{w^3}+\delta\alpha\frac{\vec{w}}{w}}\right]=0
\label{eq:MasterEquation}
\end{equation}
We take the limit of small deformation, expressed as $\frac{\partial u_x}{\partial x \hfill},\frac{\partial u_x}{\partial y \hfill},\frac{\partial u_y}{\partial y \hfill},\frac{\partial u_y}{\partial x \hfill}\ll 1$ with $\vec u = u_x\vec e_x+u_y\vec e_y$. With the following notation, $\tilde \alpha = \frac{{c^*}^2 \alpha}{\tilde Y}$, $\nu=\delta \frac{w^*}{c^*}$ and $\chi=\frac{\tilde{\alpha}-1}{2}$, the differential equation followed by $\vec u$ is:
\begin{align}
\frac{\partial^2 u_x}{\partial x^2 }+\nu\chi\frac{\partial^2 u_x}{\partial y^2 }&=0 \\
\frac{\partial^2 u_y}{\partial x^2 }+\frac{\nu}{\chi}\frac{\partial^2 u_y}{\partial y^2 }&=0
\end{align}
We now want to impose a deformation similar to what is observed experimentally in the plastic events.The displacement field of a plastic event is characterized by non-zero value of vorticity and deviatoric strain, but also by a vanishingly small shear strain and divergence. In the following we will approximate the deviatoric strain by $\frac{\partial u_x}{\partial x\hfill}-\frac{\partial u_y}{\partial y\hfill}$. We consider an event zone of size $d$ undergoing a vorticity $\omega_0$ and a deviatoric strain $\varepsilon_{d_0}$, with $\varepsilon_{d_0}\leqslant0$ because the fabric is pulled experimentally in the $y$ direction, and $\omega_0>0$ for a positive event and $\omega_0<0$ for a negative event. We therefore have the following boundary conditions around the event zone:
\begin{align}
\frac{\partial u_x}{\partial x\hfill}-\frac{\partial u_y}{\partial y\hfill}&=\varepsilon_{d_0} \\
\frac{\partial u_y}{\partial x\hfill}-\frac{\partial u_x}{\partial y\hfill}&=\omega_0 \\
\frac{\partial u_x}{\partial x\hfill}+\frac{\partial u_y}{\partial y\hfill}&=0 \\
\frac{\partial u_y}{\partial x\hfill}-\frac{\partial u_x}{\partial y\hfill}&=0
\end{align}
To solve analytically the equations for $u_x$ and $u_y$, we can use an approximate version of the initial conditions:
\begin{align}
u_x(x=\pm d,y=0)&=\pm\varepsilon_{d_0} d \\
u_x(x=0,y=\pm d)&=\mp\omega_0 d \\
u_y(x=\pm d,y=0)&=\pm\omega_0 d \\
u_y(x=0,y=\pm d)&=\mp\varepsilon_{d_0} d
\end{align}
The analytical solution with a vanishing displacement far from the event zone can then be found and reads:
\begin{align}
u_x(x,y)&={d}^2\frac{x \varepsilon_{d_0} \nu\chi-y \omega_0}{x^2\nu\chi+y^2} \\
u_y(x,y)&={d}^2\frac{x \omega_0 \frac{\nu}{\chi}-y\varepsilon_{d_0}}{x^2\frac{\nu}{\chi}+y^2} 
\end{align}
The constraint on the conservation of yarn length, written in the small deformation limit and used to evaluate the Lagrange multiplier $\alpha$, imposes:
\begin{equation}
\int\limits^{+\infty}_{-\infty}\mathrm{d}x\mathrm{d}y\left({ \frac{\partial u_x}{\partial x\hfill}+\nu \frac{\partial u_y}{\partial y\hfill}}\right)=0
\end{equation}
However, the symmetry of the displacement field induces that this condition is true whatever the value of  $\alpha$. To simplify the ensuing analysis, we choose $\tilde \alpha =2$ (or $\chi=1$). We want to observe the deformation in the stitch network, so we choose the asymmetric parameter to be equal to 1: $\nu=1$. 
With those parameters, the resulting vorticity and deviatoric fields simply write: 
\begin{align}
\omega(x,y)=d^2\varepsilon_{d_0}\frac{x y}{{(x^2+y^2)}^2}\\
\varepsilon_d(x,y)=d^2\omega_0\frac{x y }{{(x^2+y^2)}^2}
\end{align}
or in polar coordinate:
\begin{align}
\omega(r,\theta)=\frac{d^2\varepsilon_{d_0}}{ 2 r^2}\sin(2\theta) \\
\varepsilon_d(r,\theta)=\frac{d^2\omega_0}{ 2 r^2}\sin(2\theta)
\end{align} 
These formula are valid for $r>d$. 
This derivation recovers two properties measured experimentally. Deformation, in terms of vorticity and deviatoric strain, of the surroundings of a plastic event shows maxima in the directions $\frac{\pi}{4}[\frac{\pi}{2}]$, exactly what is found in the correlation function of vorticity evaluated experimentally. 
Furthermore, surprisingly, the vorticity field $\omega$ is proportional to the deviatoric strain of the event, and inversely, the deviatoric field is proportional to the vorticity of the event. This property corroborates the measured strong correlation between the deviatoric and vorticity field within the events. 

Considering asymmetric coefficients (for example $\chi\neq1$, $\nu<1$, $x_0\neq y_0$) only slightly shift the maxima and we do not recover an exact proportionality between $\omega$ and $\varepsilon_{d_0}$ and $\varepsilon_d$ and $\omega_0$.